\documentclass[letter]{aa}

\usepackage{graphicx}
\usepackage{txfonts}
\usepackage{lipsum}
\usepackage{subcaption}         
\usepackage{lscape}             
\usepackage{placeins}           
\usepackage{amsmath}
\usepackage{amssymb}
\usepackage{hyperref}
\usepackage{scalerel}
                                
\begin{document}
\makeatletter
\let\do@linenumbers\relax
\let\linenumbers\relax
\let\modulolinenumbers\relax
\let\switchlinenumbers\relax
\let\runninglinenumbers\relax
\let\pagewiselinenumbers\relax
\let\resetlinenumber\relax
\makeatother

   \title{Separating halo and disk stars in galaxies with Fuzzy Set Theory}



   \author{Amit Mondal\inst{1}
        \and Biswajit Pandey\inst{1}
        }

   \institute{Department of Physics, Visva-Bharati University, Santiniketan, 731235, West Bengal, India\\
             \email{amitmondal.bwn95@gmail.com, biswap@visva-bharati.ac.in} \\ }

   \date{Received September 30, 20XX}

 
 \abstract
  {Disk and halo stars are generally classified using several conventional methods, such as the Toomre diagram, sharp cuts in metallicity ([Fe/H]), vertical distance ($\left|Z\right|$) from the Galactic plane, or thresholds on the orbital circularity parameter ($\epsilon$). However, all these methods rely on hard selection cuts, which either contaminate samples when relaxed or exclude genuine members when applied too strictly, leading to uncertain and biased classifications.}
  {We aim to develop a more reliable and flexible approach to classify disk and halo stars in galaxies by applying fuzzy set theory, which can overcome the limitations of traditional hard-cut selection methods.}
  {We propose a fuzzy set based classification method to separate disk and halo stars in a galaxy. As a case study, we analyze one of the Milky Way/M31-like galaxies in the TNG50 catalogue. In this approach, we select multiple stellar properties and characterize their variations between disk and halo stars to construct accurate membership functions. These functions are then combined to assign each star a membership degree corresponding to its galactic component.}
  {Our fuzzy set approach provides a more realistic distinction between the disk and the halo stars. This method effectively reduces contamination and recovers genuine members that are often excluded by rigid selection criteria.}
  {The fuzzy set theory framework offers a robust alternative to conventional hard-cut methods, enabling more accurate and physically meaningful separation of stellar populations in galaxies.}


   \maketitle

\section{Introduction}
\label{sec:intro}
A galaxy is composed of a few overlapping stellar components. Galaxies like the Milky Way are composed of a bulge, a thin disk, a thick disk and an extended stellar halo \citep{gilmore83, wyse86, helmi08, bland16}. Broadly, these components can be grouped into the disk and the stellar halo. Different components encode the record of formation and assembly history of the galaxy \citep{searle78, freeman02, bullok05}. Disentangling these components is essential to reconstruct formation pathways from both surveys and simulations \citep{carollo07, ivezic08, johnston08, font11, pillepich15}, but their overlapping spatial, chemical, and kinematic distributions complicate simple, hard-edged separations.  

In observational studies, the Toomre diagram is one of the most widely used tools for distinguishing between disk and halo stars using kinematic criteria. It provides a clear visual diagnostic by plotting the rotational velocity of a star relative to the Local Standard of Rest (LSR) against the combined radial and vertical velocity components, and separates the kinematically cold disk population from the dynamically hot stellar halo \citep{venn04}. Subsequent analyses have shown that the velocity distribution of the thick disk extends toward higher velocities, producing an overlap with the halo regime \citep{schonrich09}. It indicated that adopting a higher velocity threshold is necessary to minimize contamination when identifying halo stars \citep{bonaca17}.
In \citet{gaia18}, another technique was introduced to separate halo stars from disk stars by applying a kinematic cut based on the transverse velocity ($V_T$). Typically, thin-disk stars exhibit $V_T < 40$ km/s, thick-disk stars have $60 \lesssim V_T \lesssim 150$ km/s, while halo stars generally show $V_T > 200$ km/s.

Another commonly used approach to separate halo stars from disk stars is to apply a sharp cut on the vertical distance from the Galactic plane. In a right-handed Cartesian coordinate system centered at the Galactic center, the x-axis points from the Sun toward the Galactic center, the y-axis follows the direction of Galactic rotation, and the z-axis points toward the North Galactic Pole. Following \citet{yang19}, a cut of $\left|Z\right| > 5~\mathrm{kpc}$ is often adopted, thereby excluding stars within 5 kpc above or below the Galactic plane. This effectively removes most disk stars, yielding a relatively clean halo sample. Similarly, \citet{wu22a} applied the same $\left|Z\right|$ criterion in combination with a metallicity cut of $[\mathrm{Fe}/\mathrm{H}] < -0.5~\mathrm{dex}$ to further reduce disk contamination. Another study by \citet{wu22b} adopted a slightly less stringent vertical cut of $\left|Z\right| > 2~\mathrm{kpc}$ together with the same metallicity threshold to distinguish halo stars from those in the disk. The combination of kinematic and metallicity criteria is also used to distinguish halo stars in observational data \citep{carollo07, morrison09}.

The previously discussed methods are typically applied to observational data, whereas simulations often employ a different approach. In simulations, the orbital characteristics of stars within a galaxy are described using the circularity parameter $(\epsilon)$, defined by \citet{abadi03}.  
Stellar orbits with $\epsilon \geq 0.8$ \citep{zolotov09} are generally regarded as circular, corresponding to stars in rotational motion and therefore classified as disk stars. To extract the halo component, \citet{zhu22} selected stars with $\epsilon < 0.5$ and $r > 3.5~\mathrm{kpc}$. This criterion effectively removes most stars associated with the disk and bulge components, yielding a clean sample of halo stars.

All the techniques commonly used to separate stars belonging to different galactic components suffer from a major limitation, the applied selection cuts are typically crisp or hard boundaries. When these cuts are relaxed, the resulting samples become contaminated by stars from other components, whereas applying overly strict cuts leads to the loss of stars that may genuinely belong to the intended population. Consequently, the classification becomes uncertain and potentially biased.

Fuzzy set theory was first introduced by \citet{zadeh65} and has since found widespread applications across diverse disciplines. It has been used in decision-making and industrial automation \citep{zadeh73}, control systems \citep{lugli16, Lopatin18}, image processing \citep{rosenfeld79} and pattern recognition \citep{rosenfeld84, bezdek81}, and robotics \citep{wakileh88}. The theory provides a flexible mathematical framework to represent and manipulate vagueness or uncertainty associated with systems that have imprecise or overlapping boundaries.

Fuzzy set theory has found several applications in astronomy and cosmology, particularly in problems involving uncertain or overlapping classifications. \citet{spiekermann92} introduced a fully automated morphological classification system for faint galaxies based on fuzzy algebra. Later, \citet{mahonen00} developed a fuzzy reasoning-based classifier for star-galaxy separation, which demonstrated reliable performance on real observational data and offered a promising alternative to neural network-based approaches. Extending the use of fuzzy logic to galaxy properties, \citet{coppa11} applied an unsupervised fuzzy partition clustering algorithm to the principal components obtained from a PCA analysis to investigate the bimodality in galaxy colours. More recently, \citet{pandey20} proposed a fuzzy set-based framework for galaxy colour classification that effectively addresses the uncertainties introduced by sharp boundaries in traditional colour cuts.

In this letter, we introduce a method based on the concept of fuzzy set theory, which allows for a more reliable and flexible classification of halo and disk stars. The method requires selecting multiple stellar properties, together with their known variations between disk and halo stars, in order to build accurate membership functions. These membership functions are combined to assign individual stars to their most likely Galactic component, namely the disk or the halo. The strength of our approach lies in its ability to incorporate multiple stellar properties, each reflecting differences between disk and halo populations within one coherent framework for constructing membership functions.

The letter is organized as follows. \autoref{sec:data} provides a description of the simulation data, \autoref{sec:method} outlines the methodology, and \autoref{sec:results_conclusion} presents the main results and conclusions.






\section{DATA} 
\label{sec:data}
The IllustrisTNG simulation suite \citep{nelson18, nelson19, springel18, pillepich18, marinacci18, naiman18} provides a state of the art framework for modeling galaxy formation and evolution within a cosmological context. Built on the moving-mesh code AREPO \citep{springel10, weinberger20}, it captures the interplay between dark matter, gas, and stars, including key processes such as gas dynamics, star formation, feedback, and black hole growth. TNG comprises three volumes TNG50, TNG100, and TNG300 at multiple resolution levels, all starting at $z = 127$ and adopting Planck 2015 cosmology \citep{planck16}.

For this study, we use the $z=0$ snapshot of the high-resolution TNG50-1 simulation, which follows matter in a $51.7~\mathrm{cMpc}$ box, resolving baryons with $8.5\times10^4~M_\odot$, dark matter with $4.5\times10^5~M_\odot$, and star-forming ISM gas at $\sim 100$--$140~\mathrm{pc}$ \citep{pillepich19}. We analyze the TNG50 MW/M31-like galaxy catalogue \citep{pillepich24}, which contains 198 galaxies at $z=0$.
In our analysis, we use the \textit{RotatedCoordinates} $(x, y, z)$, \textit{RotatedVelocities} $(v_x, v_y, v_z)$, and \textit{GFM\_Metals} (H and Fe abundances) of the stellar particles of a galaxy with SubhaloID 400974, which is one of the 198 galaxies from the TNG50 MW/M31-like galaxy catalogue. A detailed description of these quantities can be found at \url{https://www.tng-project.org/data/milkyway+andromeda}.

\section{Method of analysis}
\label{sec:method}
\subsection{Fuzzy sets: definitions and properties}
\citet{zadeh65} first introduced the idea of fuzzy sets to describe concepts that do not have sharp boundaries. If F is a fuzzy set that belongs to the Universal set X, it is specified by a membership function $\mu_{\scaleto{F}{3.5pt}}:X \rightarrow{}[0,1]$ and it can be represented as,
\begin{equation}
  F=\big{\{}\,(\,x,\,\mu_{\scaleto{F}{3.5pt}}(x)\,) \,\,|\,\,x \in X\,\big{\}} 
\end{equation}
The value of $\mu_{\scaleto{F}{3.5pt}}(x)$ represents the degree to which x belongs to the fuzzy set F. $\mu_{\scaleto{F}{3.5pt}}(x)=1$ denotes full membership and $\mu_{\scaleto{F}{3.5pt}}(x)=0$ denotes complete non-membership, and intermediate values indicate partial membership. Importantly, the membership degree does not represent the probability that an element belongs to a fuzzy set. Instead, membership functions describe the possibility of an element being part of a given fuzzy set. Unlike likelihood functions in probability theory, fuzzy membership functions do not need to be normalized. Fuzzy set theory is based on the idea of possibility rather than probability. Although both frameworks deal with uncertainty, probability values must sum to 1, whereas fuzzy membership degrees have no such requirement.
Because elements can belong to a set to varying degrees, fuzzy sets are naturally suited for representing vague or uncertain categories. \\

The membership functions may take many shapes depending on their context of use. The commonly known shapes are triangular, trapezoidal, Gaussian, sigmoidal, etc. Basic set operations hold naturally using membership based rules. Suppose we have two fuzzy sets F and G. The membership function of F$\cap$G is $\mu_{\scaleto{F\cap G}{3.5pt}}(x) = min\{ \mu_{\scaleto{F}{3.5pt}}(x), \mu_{\scaleto{G}{3.5pt}}(x)\}$. And the membership function of F$\cup$G is $\mu_{\scaleto{F\cup G}{3.5pt}}(x) = max\{ \mu_{\scaleto{F}{3.5pt}}(x), \mu_{\scaleto{G}{3.5pt}}(x)\}$. The membership function of the complement of F is $\mu_{{\scaleto{F}{3.5pt}}^{\scaleto{c}{3pt}}}(x) = 1-\mu_{\scaleto{F}{3.5pt}}(x)$. Thus, a fuzzy set provides a simple and flexible way to represent partial membership and uncertainty, making it well suited for analyses in which category boundaries are not sharply defined.

\subsection{Fuzzy membership functions}

Usually, stars in a galaxy are split into disk or halo categories by applying crisp cuts. For example, a sharp threshold in the circularity parameter or kinematic boundary from a Toomre diagram. This approach forces every star into one class, even when its properties lie near the dividing line. So, there is an uncertainty involved in the decision of labelling a star as either disk or halo whenever the property reaches the neighborhood of the precisely defined border.\\
Fuzzy set theory allows a star to belong partly to more than one group by assigning a membership value. So, it captures the uncertainty when the boundaries are not precise. 
In our analysis, we use stellar data from a galaxy in the TNG50 MW/M31 catalogue to construct fuzzy classifications of disk and halo stars. For each stellar property that shows distinct behaviour between the disk and halo, we incorporate that property into our analysis by defining a corresponding fuzzy set. Each fuzzy set assigns to each star a membership degree between 0 and 1 that quantifies how strongly the star resembles the disk according to that property. These memberships can be used individually or combined to study the gradual disk-halo transition.

First, we define fuzzy sets corresponding to the diskness of stars in the galaxy based on an observable y. The mathematical form of the fuzzy set corresponding to this stellar property is
\begin{equation}
 D_y = \{y,\, \mu_{\scaleto{y}{3.5pt}}^{\scaleto{D}{3.5pt}}(y) \mid y \in Y\},    
\end{equation}
where, $Y$ denote the universal set corresponding to the stellar property of all stars in the galaxy.
There is no sharp boundary separating the disk and halo stellar populations, the transition between them is gradual. We adopt sigmoidal membership functions to capture this gradual transition. Depending on the behaviour of a given stellar property y, the form of the sigmoidal membership function changes accordingly. If higher values of y correspond to increasingly disk-like characteristics, the sigmoidal membership function is 
\begin{equation}
\mu_{\scaleto{y}{3.5pt}}^{\scaleto{D}{3.5pt}}(y)=\frac{1}{1+e^{-a(y-c)}}.  
\end{equation}
Here, $a$ is the steepness, and $c$ is the midpoint ($\mu_{\scaleto{y}{3.5pt}}^{\scaleto{D}{3.5pt}}(y) = 0.5$) of the sigmoid curve.
Conversely, if the relationship is reversed, meaning lower values of y indicate more disk-like behaviour, then the membership function takes the form
\begin{equation}
\mu_{\scaleto{y}{3.5pt}}^{\scaleto{D}{3.5pt}}(y)=\frac{1}{1+e^{a(y-c)}}. 
\end{equation}

A larger ``disk membership'' is associated with a smaller ``halo membership'' and vice versa. Thus, we define a fuzzy set $H_y$ corresponding to the haloness for stars for the property $y$, as the fuzzy complement of the disk set ($H_y$). The membership function $\mu_{\scaleto{y}{3.5pt}}^{\scaleto{H}{3.5pt}}(y)$ of the fuzzy set $H_y$ can then be written as, $\mu_{\scaleto{y}{3.5pt}}^{\scaleto{H}{3.5pt}}(y) = 1 - \mu_{\scaleto{y}{3.5pt}}^{\scaleto{D}{3.5pt}}(y)$.

To obtain the overall disk membership when multiple stellar properties are used, we combine all individual disk membership values using the fuzzy minimum operator, 
\begin{equation}
\mu_{\scaleto{disk}{3.5pt}}(y) =\min \Big\{ \mu^{\scaleto{D}{3.5pt}}_{\scaleto{y_1}{3.5pt}}(y_1), \,\mu^{\scaleto{D}{3.5pt}}_{\scaleto{y_2}{3.5pt}}(y_2), \dots \,\mu^{\scaleto{D}{3.5pt}}_{\scaleto{y_n}{3.5pt}}(y_n) \Big\}.
\label{eq:disk_memb}
\end{equation}
Here, $y_1$, $y_2$,... $y_n$ denote the stellar properties selected for distinguishing disk stars from halo stars.
And the overall halo membership functions can be written as
\begin{equation}
\mu_{\scaleto{halo}{3.5pt}}(y) = 1 - \mu_{\scaleto{disk}{3.5pt}}(y),
\label{eq:halo_memb}
\end{equation}
which is the complement of the overall disk membership function.\\

The significance of this construction is that a star is considered strongly disk-like only if it simultaneously satisfies all property-based criteria. The minimum operator ensures that the weakest evidence among the property-based disk memberships determines the final disk membership. This conservative approach prevents a star from being assigned a high disk membership unless all of its properties are consistent with disk characteristics. Conversely, the halo membership function reflects the degree to which a star fails to satisfy the combined disk criteria, naturally capturing the transition between disk and halo populations.

\begin{table}[h]  
\centering
\scriptsize
\setlength{\tabcolsep}{1.5pt}
\caption{Fuzzy disk membership functions for stellar properties.}
\label{tab:membership}
\begin{tabular}{lcc}
\hline
Property ($y_i$) & Fuzzy set ($D_{y_i}$) & Membership function ($\mu_{\scaleto{y_i}{3.5pt}}^{\scaleto{D}{3.5pt}}(y_i)$)\\
\hline
{[Fe/H]}        & $\{[\mathrm{Fe/H}],\, \mu_{\scaleto{[Fe/H]}{3.5pt}}^{\scaleto{D}{3.5pt}}([\mathrm{Fe/H]}) \mid [\mathrm{Fe/H]} \in Y_1\}$ & $(1+e^{-a_1([{\mathrm Fe/H}]-c_1)})^{-1}$  \\

$\varepsilon$   & $\{\varepsilon,\, \mu_{\scaleto{\varepsilon}{3.5pt}}^{\scaleto{D}{3.5pt}}(\varepsilon) \mid \varepsilon \in Y_2\}$ & $(1+e^{-a_2(\varepsilon-c_2)})^{-1}$  \\

$v_\phi$        & $ \{v_\phi,\, \mu_{\scaleto{v_\phi}{3.5pt}}^{\scaleto{D}{3.5pt}}(v_\phi) \mid v_\phi \in Y_3\}$ & $(1+e^{-a_3(v_\phi-c_3)})^{-1}$\\

$v_r$           & $ \{v_r,\, \mu_{\scaleto{v_r}{3.5pt}}^{\scaleto{D}{3.5pt}}(v_r) \mid v_r \in Y_4\}$ & $(1+e^{a_4(v_r-c_4)})^{-1}$ \\

$v_z$           & $ \{v_z,\, \mu_{\scaleto{v_z}{3.5pt}}^{\scaleto{D}{3.5pt}}(v_z) \mid v_z \in Y_5\}$ & $(1+e^{a_5(v_z-c_5)})^{-1}$  \\

\hline
\end{tabular}
\end{table}

\begin{figure*}[h!]
   \centering
   \includegraphics[width=\hsize]{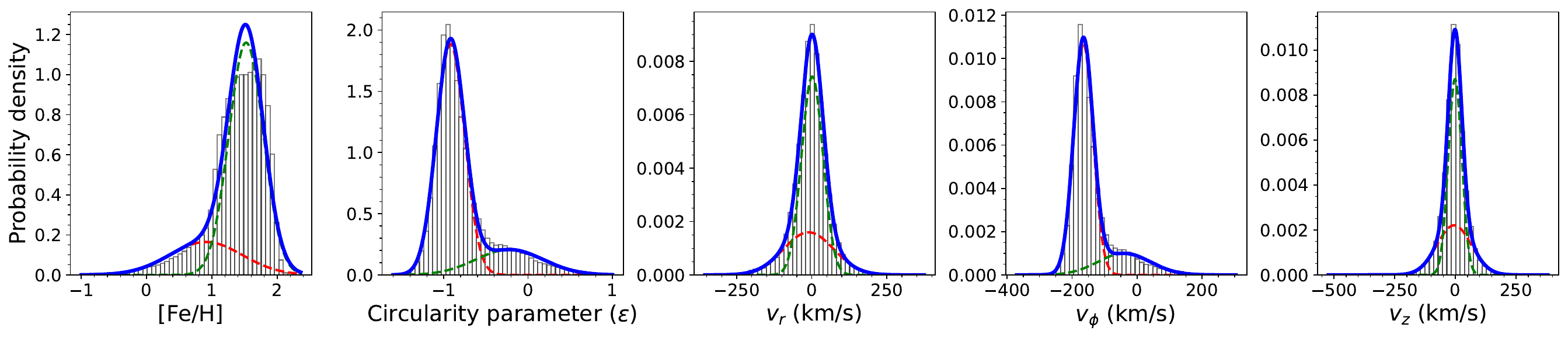}
      \caption{Probability density distributions of stellar properties for all stars in a galaxy (Subhalo ID 400974) in the TNG50 MW/M31 catalogue. The gray histograms show the simulated data. The red and green curves represent the two Gaussian components obtained from the double-Gaussian fits, while the blue curve denotes their sum (the total model). Each panel corresponds to one property: metallicity [Fe/H], circularity parameter $\varepsilon$, and velocity components $v_\phi$, $v_r$, and $v_z$. }
         \label{fig:gmm}
   \end{figure*}

\section{Results and conclusions}
\label{sec:results_conclusion}
We first exclude bulge stars by applying a radial cut of $r > 3.5$ kpc to the galaxy with Subhalo ID 400974. In our analysis, we aim to use the stellar properties [Fe/H], the circularity parameter ($\epsilon$), and the velocity components ($v_\phi$, $v_r$, $v_z$) to construct the individual diskness fuzzy sets and their corresponding fuzzy disk membership functions. The diskness fuzzy sets for the five physical properties along with the associated membership functions are provided in \autoref{tab:membership}. Estimating these membership functions requires determining the parameters $a_i$ and $c_i$ for the five properties. 

To obtain these parameters, we model the distribution of each property using a double Gaussian. Specifically, we employ the \texttt{scikit-learn} package in \texttt{Python} \citep{pedregosa11} to fit a two-component Gaussian Mixture Model (GMM) to the distributions of metallicity ([Fe/H]), circularity parameter ($\varepsilon$), and the velocity components ($v_\phi$, $v_r$, $v_z$). The fitted GMMs are shown in \autoref{fig:gmm}. 

The value of $c_i$ is defined as the intersection point of the two Gaussian components, representing the location where the disk-like and halo-like contributions are equal. Generally the two Gaussian components produce two intersection points typically one near the tail of the distributions and the other between the means of the components. However, for the radial velocity component ($v_r$) and the vertical velocity component ($v_z$), both intersection points occur in a symmetrical position around the means of the two Gaussians, as shown in \autoref{fig:gmm}, which makes it impossible to reliably identify a physically meaningful $c$. Consequently, we exclude these two properties and proceed using the remaining properties: [Fe/H], the circularity parameter ($\epsilon$), and the azimuthal velocity component ($v_\phi$) for the rest of our analysis. It may be noted that there is a gradual transition in population on both sides of the intersection of the two Gaussians for all three selected stellar properties ([Fe/H], $\varepsilon$, $v_{\phi}$) (\autoref{fig:gmm}). A hard cut at the intersection point of the two Gaussian distributions would classify the stellar populations into two distinct components. However, any value of a given property lying close to this boundary should be considered as having comparable probabilities of belonging to either component. The uncertainty is maximum near this intersection, yet it is often ignored when adopting such a hard-cut classification. Fuzzy sets avoid these hard cuts and efficiently capture the gradual transition between the disk and the halo population of stars.    
 
\begin{figure}[h!]
   \centering
   \includegraphics[width=\hsize]{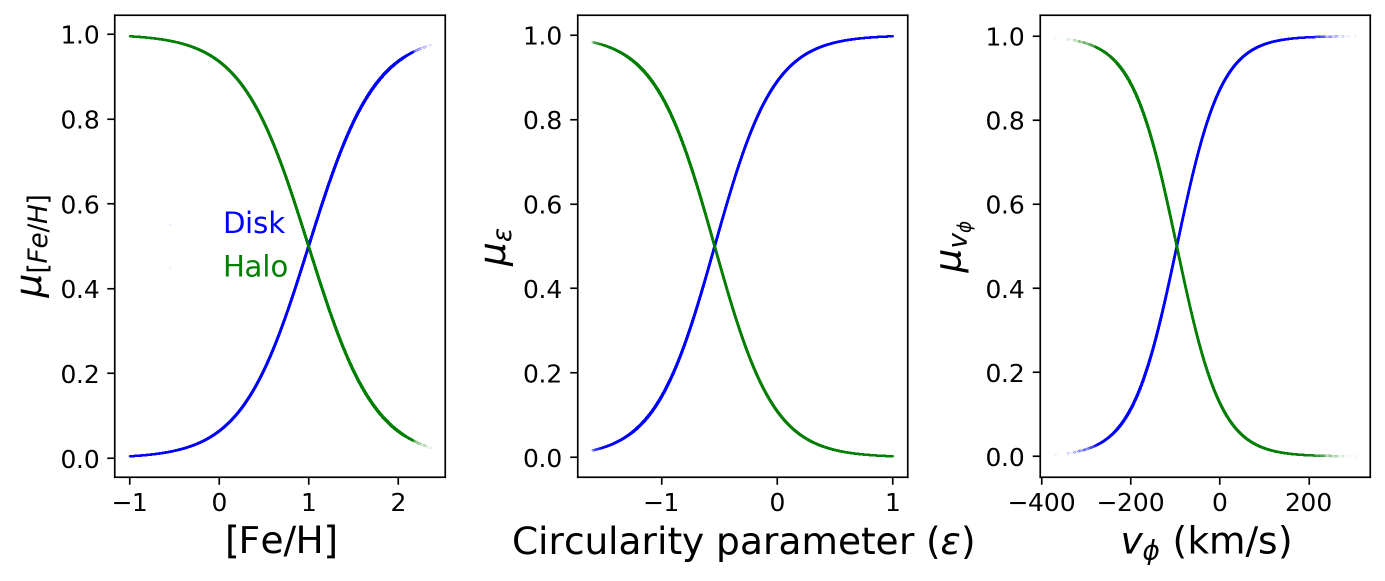}
      \caption{Different panels show the disk membership functions (blue) and halo membership functions (green) constructed for the stellar properties: metallicity [Fe/H], circularity parameter $\varepsilon$, and azimuthal velocity component ($v_{\phi}$). }
         \label{fig:memb_func}
   \end{figure}

The parameter $a_i$ controls how rapidly the sigmoidal membership function changes from low to high values. Although the disk to halo transition is gradual, $a_i$ must be chosen so that the function varies meaningfully across the region where the two stellar populations overlap, while still maintaining a smooth transition. To estimate the steepness of a sigmoid $a_i$, we use the width of its central transition region, the x-interval over which the curve rises from 10$\%$ to 90$\%$ of its maximum value. Using this estimate, we obtain $a_i=\frac{2\ln(9)}{2((\sigma_1)_i + (\sigma_2)_i)}$ where $(\sigma_1)_i$ and $(\sigma_2)_i$ are the standard deviations of the two Gaussian components corresponding to the $i^{th}$ stellar property. The resulting values of the membership function parameters are:  ($a_1 = 2.679$, $c_1 = 0.998$), ($a_2 = 3.875$, $c_2 = -0.542$), and ($a_3 = 0.020$, $c_3 = -96.998$). The disk and halo membership functions for the three stellar properties are displayed in \autoref{fig:memb_func}. In all three panels, the blue curves represent the disk membership functions, while the green curves show the corresponding halo membership functions.

By combining all three disk membership functions using the fuzzy minimum operator mentioned in \autoref{eq:disk_memb}, we obtain a unified disk membership function that accounts for the collective information from [Fe/H], $\epsilon$, and $v_{\phi}$. Stars are classified as disk when $\mu_{\scaleto{disk}{3.5pt}}(y)$ exceeds $\mu_{\scaleto{halo}{3.5pt}}(y)$, and as halo when $\mu_{\scaleto{halo}{3.5pt}}(y)$ exceeds $\mu_{\scaleto{disk}{3.5pt}}(y)$. This combined membership function provides a robust and reliable classification of stars into disk and halo components by incorporating both the kinematic and chemical properties simultaneously.

As demonstrated by the disk-halo transition presented in this work, astronomical datasets often lack sharp boundaries, making an accurate and contamination-free classification indispensable for reliable scientific interpretation. In future, we plan to extend this fuzzy classification approach to major stellar surveys like Gaia \citep{gaia23}, APOGEE \citep{majewski17}, GALAH \citep{silva15}, WEAVE \citep{jin24} and also to simulation datasets in a fully data-driven manner. We conclude that fuzzy set theory provides a powerful, reliable and flexible tool for tackling complex, high-dimensional datasets in Astrophysics and Cosmology.

\begin{acknowledgements}

AM acknowledges UGC, Government of India for support through a Junior Research Fellowship. BP acknowledges IUCAA, Pune, for providing support through the associateship programme. 

\end{acknowledgements}


\bibliographystyle{aa} 
\bibliography{ref} 

\end{document}